# Error Model in Single Board Computer based Phasor Measurement Units

Carlo Guarnieri Calò Carducci, *Member, IEEE*, Gianluca Lipari, *Member, IEEE*, Nicola Giaquinto, *Member, IEEE*, Ferdinanda Ponci, *Member, IEEE* and Antonello Monti, *Member, IEEE*

*Abstract*—Phasor Measurement Units (PMUs) are measurement devices long used in transmission systems and today even more essential for a proper monitoring of distribution grids. The expected massive penetration of distributed energy resources (DERs) is slowly taking place, carrying along a new set of challenges that put to test traditional instruments and requiring more performance and flexibility to adapt to this evolving scenario. Cheap devices based on single board computer (SBC) are proving to be a valid alternative to traditional PMU architectures, able to combine together high-performance, great versatility and low-cost. However, such devices lack a proper modeling of their measurement errors that conversely would be extremely useful for improving their design and evaluate their performance in accordance with the relevant standards. The paper intends to fill this gap by discussing the error sources and their effects on the observed signals. An analysis of error statistics is presented, in order to give a more complete metrological characterization.

## I. Introduction

From their first use at the beginning of 90s till the current date, PMUs have undergone a great mutation. Their first adoption in transmission networks (TN) aimed to detect possible oscillation between generators and describe the power flow across the network [1]. Used to monitor a unidirectional power flow in high inertia grids, early PMU prototypes [2] were low-rate data acquisition systems (DAQs) with simplified demodulation schemes borrowed from the radio signal domain to work in quasi-static conditions around the main phasor frequency. With the increasing penetration of distributed energy resources (DER), the current distribution networks (DN) are facing a growing bidirectional power flow that introduces additional dynamics to the network, resulting in enhanced short-time power fluctuations and faster frequency variations. Furthermore, the much smaller phase differences as a result of the reduced line lengths at the distribution scale and the more complex network topology demand a greater number of measuring devices and at the same time improved measurement accuracy [3]. Considering that in the last decade the number of deployed PMUs across U.S. and Canada has grown by a factor of more than ten, from 200 research-grade PMUs to more than 2500 production-grade [4], it is easy to understand how traditional PMUs would constitute an unbearable burden for the distribution system operators (DSO).

As a consequence, a great variety of low-cost solutions are emerging in the recent years, ranging from high-performance FPGA [5] boards to open platform solutions like the OpenPMU project [6], which is based on National Instruments NI-DAQ data acquisition devices. Despite their stated cost around 1000$ results much lower than traditional instruments, still it is too high for a realistic deployment in DN. Recent researches [7], however, are focusing on PMU architectures based on system on chip (SoC), embedded solutions which are proving to provide acceptable performance while further reducing the hardware cost by one order of magnitude. The SoC category can be further split in single board microcontroller and single board computer (SBC), consequently leading to different PMU design approaches. Microcontrollers are very efficient when performing specific tasks, they support hardware interrupts and real-time events, but they require a dedicated firmware. Conversely, the operating system (OS) running on an SBC can potentially provide greater flexibility and versatility in adapting to a rapidly evolving scenario. However, this comes at the cost of losing real-time control on the interrupts. In both cases, the main accuracy challenges arising from the use of cheap SoC consist in the synchronization of the sampling base with the pulse-per-second (PPS) signal, in the stability of the sampling base itself and on the conversion stage.

C. Guarnieri Calò Carducci, G. Lipari, F. Ponci, A. Monti are with the Intitute for Automation of Complex Power Systems, E.ON Energy Research Center, RWTH Aachen University, 52074 Aachen, Germany (e-mail: cguarnieri@eonerc.rwth-aachen.de; glipari@eonerc.rwth-aachen.de; fponci@eonerc.rwth-aachen.de; amonti@eonerc.rwth-aachen.de).

N. Giaquinto is with Dipartimento di Ingegneria Elettrica e dell'Informazione, Politecnico di Bari, 70125 Bari, Italy (e-mail: nicola.giaquinto@poliba.it).

Regarding the microcontroller approach, authors in [8] present a solution based on an ARM Cortex-M4 that makes use of the three embedded ADC and of an external GPS module. However, tests have been performed only in static conditions and without considering environmental effects. In [9] authors mainly improve the previous solution by proposing a PID controller connected to the PPS for resampling the acquired waveform in the correct time instants. In both cases, though, the monitoring of currents and voltages in a tri-phase system would require multiple boards.

In the SBC approach, two main strands can be identified. A first class of OS-based PMU solutions rely on the Beagle Bone Black (BBB) board, an ARM-based architecture running Linux operating system. Authors in [10] and [11] have analyzed the performance of this board with respect to the cascade execution of two state-of-the-art real-time algorithms for synchrophasors estimation, i.e. the Interpolated DFT (IpDFT) and the Taylor-Fourier Transform (TFT). Their analysis reports a processing time smaller than 20 ms, supporting its possible use in the development of low-cost instruments for large-scale grid monitoring. However, the authors also highlight the need for an ad-hoc acquisition and synchronization stage, a solution adopted in the newer version of the OpenPMU [12]. In this case, the authors implemented a GPS-disciplined ADC with the aid of an external digital PLL (DPLL) and by leveraging two programmable real-time units (PRU) embedded in the BBB processor for implementing those real-time tasks hardly handled by the operating system. This mechanism has been furtherly improved in [13] by removing the DPLL and implementing a software-based low-jitter servo clock directly inside one of the PRUs.

A second class of OS-based PMU solutions relies on the well-known Raspberry Pi board (RPi), a widely adopted multi-purpose ARM platform running Raspbian OS, a free Linux distribution optimized for its hardware. Its versatility emerges for instance from its adoption for educational purposes: for example, in conjunction with Matlab Simulink blocks, it is used to propose laboratory exercises on PMU algorithms for undergraduate students in UK and Mexico [14]. A concrete application of a RPi as fully functional PMU was introduced with the LoCo PMU [15], where the authors present a low-cost device that integrates an external DAQ board and a GPS module. The solution to the synchronization problem consists in this case in the direct triggering of the DAQ acquisition by using the PPS from the GPS module. The authors also discuss the calibration of such device over the entire signal chain, from the instruments transformer to the PMU [16]. A recent evolution of the same device relies [17] on a dedicated conversion stage rather than on a third party DAQ, on an innovative software PLL based on a Kernel module and on the use of the internal pulse width modulation block (PWM) to generate the sampling base.

However, to the best of the author's knowledge, a clear modeling from the design perspective of the measurement error generation mechanisms in SBC and of their influence on the errors specified by the standard IEEE C37.118.1 is still missing in literature and therefore discussed in this work with the aim of improving design and calibration techniques.

The paper is structured as follows. In Section II the authors describe the error mechanisms with respect to the last mentioned SBC-based PMU solution. In Section III and IV respectively the adopted experimental setup is described and the performed uncertainty analysis is detailed. Finally, results are discussed in Section V and conclusions are presented in Section VI.

## II. ERROR MODEL

The signal chain of a PMU system, as shown in Fig. 1, can be subdivided in three blocks: sensor, data acquisition system and signal processing algorithm. Given their sequential nature, errors introduced in a block inevitably propagate to the following, cumulating on the final measurement.

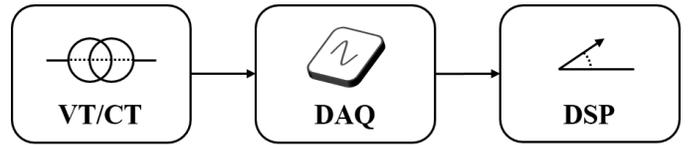

Fig. 1. PMU signal chain: voltage/current transformer (left), synchronized data acquisition system (center) and phasor estimation (right)

Errors introduced by the transducers contribute significantly to the overall uncertainty, typically in the range of the class of accuracy declared by the manufacturer. However, no much room for improvements is left since they are usually commercial solutions to be used as-is. On the other hand, a great variety of phasor estimation algorithms is already available in literature, whose comparative performance can be evaluated by means of either numerical simulation [18][19] or test bed characterization [20]. A separate discussion is required for the acquisition block, where PLL closed loop compensation techniques make very difficult, if not impossible, to clearly separate the error contributions. Conversely, in the investigated RaspberryPi-based PMU implementation discussed in [17], each error source can be clearly defined and characterized.

The proposed model of the investigated PMU data acquisition system can be subdivided into four blocks as shown in Fig. 2: the input anti-aliasing filter (AAF), the analog-to-digital converter (ADC), the time-reference module (GPS) and the single board computer (RPi). For each block, the direct effect of influence factors on the measured signals can be identified.

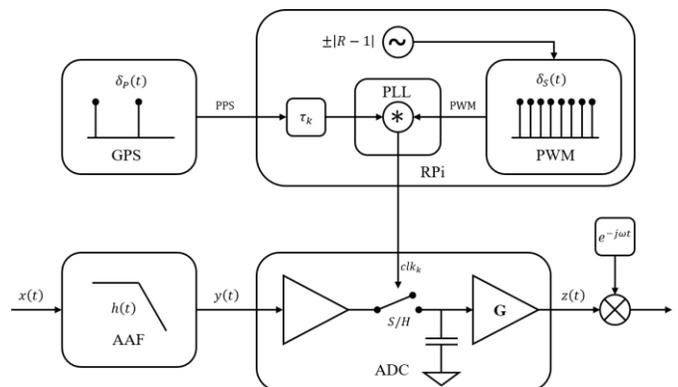

Fig. 2. Synchronized DAQ architecture of the RaspberryPi-based PMU

In the following, real signals are treated using their equivalent analytic representation, which allows for a simplification of notation and computations, and for an easy generalization of the concept of phasor from time-invariant amplitude, phase and frequency to time-variable parameters. For a real input signal

$$x(t) = A\cos(2\pi f t + \varphi) \quad (1)$$

with $A$ and $\varphi$ respectively the amplitude and phase of a periodic signal at the frequency $f$, the corresponding analytic signal is

$$x_a(t) = A e^{j\omega t + \varphi} = A e^{j\varphi} e^{j\omega t} = X e^{j\omega t} \quad (2)$$

where the time-invariant component X is called phasor. The original signal can always be recovered from the analytic signal by extracting the real part

$$x(t) = \Re\{x_a(t)\} = \Re\{X e^{j\omega t}\} . \quad (3)$$

In the case of time-variable parameters, the phasor X can be extended to the more general form of complex envelope

$$X(t) = x_a(t) e^{-j\omega t} = A(t) e^{j\varphi(t)} , \quad (4)$$

which can be obtained as baseband demodulation of the analytic signal, also known as dynamic phasor. The parameters $A(t)$ and $\varphi(t)$ now respectively identify an amplitude and a phase modulation of the carrier signal and therefore take the name of instantaneous amplitude and instantaneous phase.

The input signal $x(t)$ in (1) is first fed into the AAF, to which corresponds a convolution in the time-domain with the filter impulse response $h(t)$

$$y(t) = (x * h)(t) , \quad (5)$$

equivalently expressed in analytic form as

$$y_a(t) = X\dot{H} e^{j\omega t} = Y e^{j\omega t} \quad (6)$$

where $\dot{H} = H e^{j\varphi}$ is the Fourier transform of $h$ evaluated in a specific $\omega$ and Y is the complex envelope of the filtered signal. Hence, the filtered signal $y(t)$ is converted to digital by the ADC block, whose output can be described as a linear combination of the input via a static transfer characteristic with gain $G$ and offset $V_{OS}$. However, the offset value has no theoretical meaning in the context of phasors and practical effect for PMU operations limited to possible spectral leakage. This effect should be taken into account on case-by-case basis in relation to the adopted phasor estimation algorithm. For this reason it is not taken into account on the output-sampled signal.

The RPi is responsible for generating the sampling base for the ADC. It implements a synchronization mechanism of the time base with the time reference that is conceptually similar to a PLL, however without using any kind of close loop control. The time reference information generated by the GPS module is carried by the pulse-per-second (PPS) signal, mathematically defined by (7) as an infinite ideal pulse train $\delta_p$ with period $T$

$$\delta_p(t) = \sum_{k=-\infty}^{\infty} \delta(t - kT) , \quad (7)$$

where $\delta(t)$ is the Dirac delta function. At the same time, the hardware PWM block inside the RPi microcontroller is responsible for generating the conversion time base

$$\delta_s(t) = \sum_{n=-\infty}^{\infty} \delta(t - nRT_s) , \quad (8)$$

an infinite pulse train with period $RT_s$, where

$$R = \frac{\hat{T}_s}{T_s} = \frac{F_s}{\hat{F}_s} \quad (9)$$

is the ratio of the nominal sampling frequency $F_s$ to the actual sampling frequency $\hat{F}_s$, a deviation due to the oscillator accuracy and its drifts with temperature and age that results in a scaling of the time axis. Because of this deviation (9), the synchronism between the time base and the time reference cannot be retained with time and the two signals drift with respect to each other. The developed software PLL consists of a Kernel module that detects the interrupts generated by the PPS signal in quasi real-time and realigns the PWM time base. Therefore, equation (8) can be rewritten accordingly as

$$\delta_s(t) = \sum_{N_s} \delta(t - nRT_s) , \quad (10)$$

a finite pulse train of $N_s = T/T_s$ pulses with approximate total duration $T$, where the approximation to $N_s$ pulses holds for an absolute deviation smaller than the actual time base period

$$|R - 1| \cdot N_s < 1 . \quad (11)$$

For each PPS pulse, denoted with subscript $k$, the software PLL triggers the sequence (10) after a random delay $\tau_k$. The result is the sampling base used to drive the ADC

$$clk_k = \sum_{N_s} \delta(t - nT_s R - \tau_k)\bigg|_{t \in [k, k+1[ \cdot T} \quad (12)$$

The delay $\tau_k$ is the result of two combined random processes: the OS synchronization jitter and the PPS jitter. However, the PPS jitter of the GPS module is typically in the order of 10 ns, almost three orders of magnitude smaller than the OS synchronization jitter and thus can be considered negligible. The result of the software PLL is a reset of the combined effects of the synchronization and time base error every second, a simple mechanism that results quite effective, provided that the total error over the period does not exceed the synchronization error threshold defined by the standard IEEE C37.118.1.

The resulting sampled signal can finally be expressed as the product of (5) with the convolution of (7) and (12)

$$z(t) = y(t) G \sum_{k=-\infty}^{\infty} \sum_{N_s} \delta(t - nT_sR - kT - \tau_k), \quad (13)$$

and the corresponding analytic representation over an arbitrary PPS interval $k$ as

$$\begin{aligned} z_a(t) &= YGe^{j\omega\tau_k}e^{j\omega Rt} \\ &= Y[Ge^{j\omega\tau_k}e^{j\omega(R-1)t}]e^{j\omega t} \\ &= X[G\dot{H}e^{j\omega\tau_k}e^{j\omega(R-1)t}]e^{j\omega t} \quad (14) \\ &= \dot{\Lambda}Xe^{j\omega t} \\ &= Ze^{j\omega t} \end{aligned}$$

where

$$\dot{\Lambda} = \prod \Lambda_i \cdot e^{j\varphi_{\Lambda_i}} = \dot{\Lambda}_H \cdot \dot{\Lambda}_G \cdot \dot{\Lambda}_R \cdot \dot{\Lambda}_\tau \quad (15)$$

represents the combined response of the measuring system and the subscripts $i$ refer to the each of the previously discussed effects of $G$, $H$, $R$ and $\tau_k$.

A further multiplication of (14) by the term $e^{-j\omega t}$ allows conceptually the subsequent extraction of the baseband information contained in the system output

$$Z = \dot{\Lambda} \cdot X , \quad (16)$$

an operation that is carried out by different algorithms with variable performance and whose discussion falls outside the scope of this work.

Wherever necessary, equation (16) provides a means for compensating the effect of the measuring system. In principle, a further product of Z by the inverse of the system response $\dot{K} = 1/\dot{\Lambda}$, would restore the original phasor X

$$Z = \dot{F} \cdot X = X \iff \dot{F} = \dot{K} \cdot \dot{\Lambda} = 1 . \quad (17)$$

However, the system response $\dot{\Lambda}$ is not known and must be measured. Therefore, its true value is not known, but only its observed value $\widehat{\dot{\Lambda}}$, which can be regarded as a random variable with a given probability density function (PDF) and a standard deviation induced by the PDF of the errors. In terms of errors, this corresponds to an error on both the amplitude and the phase

$$\begin{aligned} e_\Lambda &= \widehat{\Lambda} - \Lambda \\ e_{\varphi_\Lambda} &= \widehat{\varphi}_\Lambda - \varphi_\Lambda . \end{aligned} \quad (18)$$

Consequently, the known value of the compensated response is

$$\begin{aligned} \widehat{\dot{F}} &= \widehat{\dot{\Lambda}} \cdot \dot{K} \\ &= \frac{(\Lambda + e_\Lambda) \cdot e^{j(\varphi_\Lambda + e_{\varphi_\Lambda})}}{\Lambda \cdot e^{j\varphi_\Lambda}} \quad (19) \\ &= \left(1 + \frac{e_\Lambda}{\Lambda}\right) \cdot e^{je_{\varphi_\Lambda}} \\ &= (1 + e_{r_\Lambda}) \cdot e^{je_{\varphi_\Lambda}} , \end{aligned}$$

which can be written using a more compact notation as

$$\widehat{\dot{F}} \cong e^{e_{r_\Lambda}} \cdot e^{je_{\varphi_\Lambda}} = e^{e_{r_\Lambda} + je_{\varphi_\Lambda}} . \quad (20)$$

The approximation $(1 + e_{r_\Lambda}) \approx e^{e_{r_\Lambda}}$ can be obtained for $e_{r_\Lambda} \ll 1$ by truncating the Taylor expansion $e^x = 1 + x + \frac{1}{x^2}...$ at the second term.

The errors on the observed value of $\dot{F}$ are defined with respect to its ideal values of unity magnitude and null phase

$$\begin{aligned} e_F &= \widehat{F} - 1 = e_{r_\Lambda} \\ e_{\varphi_F} &= \widehat{\varphi}_F - 0 = e_{\varphi_\Lambda} . \end{aligned} \quad (21)$$

Therefore, the observed value of the compensated output is

$$\widehat{Z} = \widehat{\dot{F}} \cdot X = [(1 + e_{r_\Lambda}) \cdot e^{je_{\varphi_\Lambda}}] \cdot X , \quad (22)$$

and the residual errors introduced by the system on the amplitude and phase of the input phasor are

$$\begin{aligned} e_Z &= |\widehat{Z}| - |X| = e_F \cdot X \\ e_{\varphi_Z} &= \widehat{\varphi}_Z - \varphi_X = e_{\varphi_F} . \end{aligned} \quad (23)$$

The uncertainty on the compensated output $\widehat{Z}$ can be defined only after assigning to all the error terms in (23) a state-of-knowledge distribution, i.e. a probability density function that represents their statistical information as random variables. Consequently, for each identified system response, $\widehat{\dot{\Lambda}}$ can be expressed in terms of true value and error as

$$\widehat{\dot{\Lambda}} \cong \widehat{\Lambda} \cdot e^{e_{r_\Lambda}} \cdot e^{j(\widehat{\varphi}_\Lambda \pm e_{\varphi_\Lambda})} , \quad (24)$$

where the true values are replaced by their expected value $\overline{\Lambda}$ and $\overline{\varphi}$ and the errors are replaced with the uncertainties $u_\Lambda$ and $u_{\varphi_\Lambda}$ associated to the standard deviation of their respective PDF. Hence, the standard uncertainties on the output amplitude and phase are

$$\begin{aligned} u(Z) &= \frac{u_\Lambda}{\Lambda} \cdot X \\ u(\varphi_Z) &= u_{\varphi_\Lambda} . \end{aligned} \quad (25)$$

where $u_\Lambda$, $u_{\varphi_\Lambda}$ are the standard deviations of the PDF of $e_\Lambda$, $e_{\varphi_\Lambda}$, respectively.

The above discussed error model can be related to the two main errors defined in the standard IEEE C37.118.1-2011 [21]. The Total Vector Error (TVE) is defined as the normalized module of the error vector between the observed and the true phasor, whereas the Frequency Error (FE) is the module of the absolute error on the estimation of a measured signal frequency. Applying (22) to the definition of the TVE allows to evaluate the effect of the measuring system response directly on the TVE

$$\text{TVE} = \left|\frac{\widehat{Z} - X}{X}\right| = |\widehat{F} - 1| . \quad (26)$$

Conversely, the effect of the system on the estimation of an unknown frequency $f$ and thus on the FE cannot be expressed in terms of phasor error, but rather with respect to the oscillator accuracy defined in (9) by the gain error $R$

$$\text{FE} = |f - \hat{f}| = \left|f - f \cdot \frac{F_s}{\hat{F}_s}\right| = f \cdot |1 - R| \quad . \quad (27)$$

The limits imposed by the standard for these two errors under steady-state conditions are 1% and 5 mHz for the TVE and FE respectively, for both M and P class PMUs.

### III. EXPERIMENTAL SETUP

All measurements described in the following are carried out using a NI USB-6356 board by National Instruments, a high-performance DAQ device with synchronous mixed-signals IO capabilities up to 1 MS/s. The analog accuracy of the device is assessed according to the procedure described in the device specifications [22]. It is evaluated for a single sample over the full range of ±10 V, within 1 °C from the auto-calibration and for a temperature change of 20 °C from the last external calibration. Calculated values are 262 ppm and 311 ppm respectively in acquisition and generation mode. The manufacturer also declares a timing resolution of 10 ns and a timing accuracy of 50 ppm of the sample rate, which results in a phase accuracy of 15.7 nrad at 50 Hz for the maximum sample rate. The overall effect of the assessed accuracy is quantifiable in a maximum total vector error (TVE) of 0.041% and 0.044% respectively in acquisition and generation mode, and a maximum frequency error (FE) of 2.5 mHz, both appropriate for conformance testing of PMU based on the standard IEEE C37.118.1.

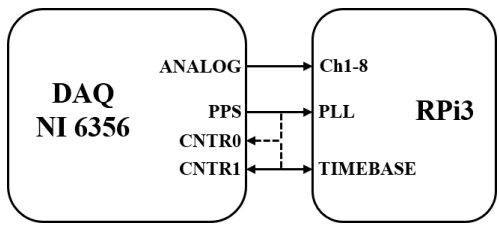

Fig. 3. Experimental setup scheme: Cntr0 is used to measure the synchronization delay of the PLL. Cntr1 is used to measure the frequency deviation R or to generate the sampling base in the ADC characterization

The device under test (DUT) is a Raspberry Pi 3 (Model B) single board computer equipped with the PMU hardware-on-top previously presented in (25). The performance of the PMU system is mainly defined by the following components:

- SoC: Broadcom BCM2837, a 64-bit quad-core ARM Cortex-A53 CPU running at 1.2 GHz
- RAM: Elpida Memory, 1 GB LowPower-DDR2 SDRAM at 400 MHz
- ADC: Texas Instruments ADS8588S, a bipolar input 16-bit simultaneous sampling SAR converter with throughput up to 200 kSPS
- GPS: Skylab SKG09BL, with 10 ns RMS timing accuracy of the PPS signal

In order to assess the repeatability of measured performance, tests are carried out over three DUT with same configuration.

For the assessment of the temperature influence on the oscillator stability, tests are performed inside an MK 53 environmental simulation chamber by BINDER, suitable for heat and cold testing in the range from -40 °C to 180 °C with a maximum temperature error over the entire range of ±2.0 K.

### IV. UNCERTAINTY ANALYSIS

In order to implement the compensation technique described at the end of Section II, it is essential to characterize first the four error sources defined in (15). The methodology approach followed for the analysis of each source is discussed for sake of clarity in separate subsections.

#### A. $\dot{\Lambda}_H$ - Response of the AAF block

The frequency response of the anti-aliasing filter can be evaluated either experimentally or analytically. However, the latter approach is widely adequate in the characterization of a simple first-order low-pass filter. In this case, given

$$H(j\omega) = \frac{1}{1 + j\omega\tau} \quad (28)$$

the AAF transfer function with a time constant $\tau = RC$, the magnitude and the phase frequency responses are

$$H \equiv |H| = \frac{1}{\sqrt{1 + \omega^2\tau^2}} \quad (29)$$
$$\varphi \equiv \angle H = -\tan^{-1}(\omega\tau) \quad .$$

Hence, by applying the uncertainty propagation formula to (29) it is possible to define the standard uncertainty of both the phase and the magnitude responses as

$$u_H = u_\tau \cdot H^3 \omega^2 \tau$$
$$u_\varphi = u_\tau \cdot H^2 \omega \quad , \quad (30)$$

where $u_\tau = \sqrt{R^2 u_C^2 + C^2 u_R^2}$ is the uncertainty of the time constant and $u_C, u_R$ are respectively the standard uncertainty associated with the tolerance of the resistor $R$ and the capacitor $C$ used in the filter. In terms of phasor notation, the effect introduced by the AAF response at a given frequency can be written as

$$\hat{\Lambda}_H = \hat{H} \cdot e^{j\hat{\varphi}}$$
$$= [1 \pm u_{r_H}] \cdot H \cdot e^{j[\varphi \pm u_\varphi]} \quad (31)$$
$$\cong H \cdot e^{u_{r_H}} \cdot e^{j[\varphi \pm u_\varphi]} \quad .$$

Considering a filter cutoff frequency $f_t$ two orders of magnitude higher than the line frequency $f$, the resulting 50 ppm magnitude attenuation is substantially negligible on the TVE, whereas the –10 mrad phase shift introduced by the filter accounts alone for 1% TVE and thus must be compensated. Similarly, if selecting passive components with standard tolerance (i.e. 10% capacitors and 1% resistors), the calculated uncertainty on the magnitude and on the phase response is 5.8 ppm and 0.58 mrad respectively. It follows that compensating only for the phase shift would result in a residual uncompensated error

$$\widehat{\Lambda}_H = H \cdot e^{\pm u_{r_H}} \cdot e^{\pm j u_\varphi} \approx e^{\pm u_{r_H} \pm j u_\varphi} \qquad (32)$$

that contributes for less than 0.06% to the TVE. If improved accuracy is required, further compensating for the magnitude attenuation and reducing the passive components tolerance by one order of magnitude (i.e. 1% capacitors and 0.1% resistors), would result in a 0.006% residual error on the TVE.

### B. $\dot{\Lambda}_G$ – Response of the ADC block

The overall accuracy and repeatability of the analog-to-digital conversion process does not only depend on the specifications of the ADC in strict-sense, but also on the influence of the analog front-end. The repeatability depends on the noise of the system and can be defined with dynamic specifications like the signal-to-noise ratio (SNR), total-harmonic-distortion (THD) and spurious-free dynamic range (SFDR), whereas the accuracy can be defined by a static characteristic with gain $G$ and offset $V_{os}$ that combines together the following effects

- actual LSB size resulting from the voltage reference
- ADC loading effects on the analog front-end
- amplification/attenuation stage used to match the sensor output swing with the ADC full-scale
- signal ground mismatches due to parasitic resistance

Gain and offset do not account for the ADC nonlinearity, which has two contributions, quantization error and integral nonlinearity (INL). Quantization error introduces a quantization noise floor with RMS value equal to $Q/\sqrt{12}$, being $Q$ the LSB size: this noise adds up in quadratic mean to the RMS system noise. Integral nonlinearity introduces spurious harmonics but, by definition, does not affect the sinusoidal signal at the fundamental frequency that is object of investigation.

The characterization of the block is carried out by measuring the ADC output $v_o$ while sweeping the channel input $v_i$ over the full-scale range (FS), and using the equation

$$v_o = G \cdot v_i + V_{os} \qquad (33)$$

where the output voltage is directly obtained from the nominal value of the voltage reference as $v_o = \text{CODE} \cdot V_{ref}/2^n$.

In order to preserve the timing alignment of generated and acquired samples, the DAQ board simultaneously generates both the analog signal and the sampling base used by the ADC. The maximum slew-rate SR that can be adopted to perform the sweep test under static conditions in a time $\tau_r$ over the range FS, can be estimated by evaluating the errors on the static characteristic introduced by the AAF block

$$\begin{aligned} e_{H,G} &= |H_{\omega_r}| - 1 \\ e_{H,os} &= -\text{SR} \cdot \Delta T = -\text{SR} \cdot \Delta\varphi \cdot \tau_r \end{aligned} \qquad (34)$$

where $\Delta T$ is time delay corresponding to the phase shift $\Delta\varphi$ introduced by the filter at the angular sweep frequency $\omega_r$, bound by the relation

$$\text{SR} = \frac{\text{FS}}{\tau_r} = \omega_r \cdot \text{FS} = \frac{\Delta\varphi}{\Delta T} \text{ FS} \quad . \qquad (35)$$

Considering a sweep duration five orders of magnitude higher than the time constant of the filter, equivalent to a slew-rate of 6.3 V/s and a sweep time of 3.2 s, the resulting 0.05 ppb gain error is largely negligible as well as the -200 µV offset error being less than one LSB of the DAQ board.

After acquiring N samples of the ADC output voltage with respect to the input voltage, the gain $G$ and the offset voltage $V_{OS}$ are determined via linear regression with OLS method by solving the problem

$$\boldsymbol{v_o} = \boldsymbol{X}\boldsymbol{\beta}, \qquad \boldsymbol{\beta} = \begin{bmatrix} V_{os} \\ G \end{bmatrix} \qquad (36)$$

where $\boldsymbol{\beta}$ is the parameters column vector, $\boldsymbol{v_o}$ is the Nx1 vector of the measured output values and $\boldsymbol{X} = [\mathbf{1} \ \boldsymbol{v_i}]$ is the Nx2 matrix of the regressors, in which the values of the input $\boldsymbol{v_i}$ are introduced. The variance of the estimator $\hat{\beta}$ can be obtained by the covariance matrix

$$\boldsymbol{K} = \frac{\text{RSS}}{N-2}(\boldsymbol{X}'\boldsymbol{X})^{-1} \cong \sigma_{\hat{\beta}}^2 \boldsymbol{I_2} , \qquad (37)$$

where RSS is the residual sum of squares and $(N-2)$ the number of degrees of freedom of the unbiased sample variance, evaluated from the residual $\boldsymbol{v_o} - \boldsymbol{X}\hat{\beta}$ assuming homoscedastic errors. Additional statistics on the inter-channels and inter-DUT parameters dispersion are obtained through simultaneous characterization of all channels of each DUT with a common input signal. If $X$ is a random variable over the sample space of the DUTs, the conditional sample variance and the total sample variance can be defined as

$$\begin{aligned} \sigma_{\hat{\beta}_X}^2 &= \mathbb{E}[\sigma^2(\hat{\beta}|X)] \\ \sigma_{\hat{\beta}}^2 &= \mathbb{E}[\sigma^2(\hat{\beta}|X)] + \sigma^2(\mathbb{E}[\hat{\beta}|X]) \end{aligned} \qquad (38)$$

and the following relation holds

$$\sigma_{\hat{\beta}} \ll \sigma_{\hat{\beta}_X} \leq \sigma_\beta , \qquad (39)$$

being $\sigma_{\hat{\beta}_X}^2$ the expected variance of $\beta$ while averaging over all values of $X$ and $\sigma_\beta^2$ the sample variance of $\beta$ from the law of total variance. In fact, $\sigma_{\hat{\beta}}$ is mainly the result of the noise of the system and of the ADC non-linearity, which is usually much smaller than the tolerances between channels of the same DUT accounted for by $\sigma_{\hat{\beta}_X}$. The additional term in $\sigma_\beta$ then accounts for the tolerance between different DUTs. Conversely, from the law of total expectation, the parameters expected value can be simply estimated as the average of all the conditional expected values

$$\bar{\beta} = \mathbb{E}[\beta] = \mathbb{E}[\mathbb{E}[\beta|X]] = \begin{bmatrix} \bar{V}_{os} \\ \bar{G} \end{bmatrix} , \qquad (40)$$

which converges to $\mathbb{E}[\beta]$ for the dominated convergence theorem. However, the effect introduced by the conversion block can be expressed with respect to the only statistical properties of the gain error, whereas the offset error can be neglected if the compensation is applied in the phasor domain rather than in time domain. In terms of phasor system response the effect of the conversion block is

$$\begin{aligned}\hat{\Lambda}_G &= \hat{G} \cdot e^{j\hat{\varphi}} \\ &= [1 \pm u_{r_G}] \cdot \bar{G} \cdot e^{j0} \cong \bar{G} \cdot e^{\pm u_{r_G}} ,\end{aligned} \quad (41)$$

where $u_{r_G}$ is the relative standard uncertainty associated to $\sigma_\beta$ and the phase term is null given the intrinsic static nature.

TABLE I
STATISTICS OF THE ADC BLOCK

|  | $\mathbb{E}[\cdot]$ | $\sigma_{\hat{\beta}}$ | $\sigma_{\beta_X}$ | $\sigma_\beta$ |
|---|---|---|---|---|
| $e_V$ (μV) | -269 | 2 | 212 | 255 |
| $e_G$ (ppm) | -4459 | 3 | 66 | 134 |

Comparative tests performed on different DUTs have confirmed (see TABLE I) experimentally what expected in (39). Results also show that the expected value of the gain would alone result in a TVE equal to 0.43 %. However, after compensation, the residual effect of the gain uncertainty would contribute to the TVE to the extent of only 0.013 %.

C. $\dot{\Lambda}_R$ – Response of the PWM block

The accuracy of the time base directly affects the frequency error FE as previously discussed in Section II, but its deleterious effects additionally manifest on the TVE in the form of a cyclostationary process whose magnitude oscillates at a frequency equal to the FE. In the discussed architecture, the PLL keeps the TVE under control by periodic synchronization of the time base with the PPS, but if correctly characterized it can be also compensated.

The time base accuracy is assessed via One Counter method by measuring the generated sampling frequency $F_s$ in the range 5-50 kHz. For each period $T_s$ of the unknown frequency, the DAQ primary counter counts the $N$ edges of the internal known time base $F_k$ (100 MHz) and the time base deviation $R$ is estimated in accordance to (9) as

$$\hat{T}_s = \frac{N}{F_k} = NT_k \quad \rightarrow \quad R = \frac{\hat{T}_s}{T_s} . \quad (42)$$

Since the error $e_R$ is proportional to the measured frequency, $N_s$ measurements are averaged for each $F_s$ in order to keep the maximum error on the mean $e_{\bar{R}}$ below 1 ppm

$$e_R = R - 1 = \frac{\hat{T}_s - T_s}{T_s} = \frac{T_k}{T_s} \quad \rightarrow \quad e_{\bar{R}} = \frac{e_R}{\sqrt{N_s}} . \quad (43)$$

Statistics over $R$ can then be calculated by performing comparative test on different DUTs and under different environmental conditions, with the purpose to assess the board-to-board dispersion $\sigma_{R_X}^2$ and the temperature dependence $\sigma_{R_T}^2$

$$\begin{aligned}\sigma_{\hat{R}}^2 &= \mathbb{E}[\mathbb{E}[\sigma^2(R|X)]|T] \\ \sigma_{R_X}^2 &= \mathbb{E}[\sigma^2(R|X)] + \sigma^2(\mathbb{E}[R|X]) \\ \sigma_{R_T}^2 &= \mathbb{E}[\sigma^2(\mathbb{E}[R|X]|T)] + \sigma^2(\mathbb{E}[\mathbb{E}[R|X]|T]) ,\end{aligned} \quad (44)$$

where $T$ is a random variable over the sample space of the operating temperature range 0-50 °C. The following relation between the estimated parameter dispersion is found to hold

$$\sigma_{\hat{R}} \leq \sigma_{R_X} \leq \sigma_{R_T} , \quad (45)$$

being the standard deviation of the estimator fixed by the measurement technique and the variation among different boards smaller than the thermal drifts. In terms of phasor notation, the response of the PWM block is

$$\begin{aligned}\hat{\Lambda}_R &= e^{j\omega(\hat{R}-1)t} \\ &= e^{j\omega([\bar{R}\pm u_R]-1)t} \\ &= e^{j\omega(\bar{e}_R\pm u_R)t}\end{aligned} \quad (46)$$

where $u_R$ is the relative standard uncertainty associated to $\sigma_{R_T}$ and the variable $t$ reflects the time dependence of the effect.

Results of the test on different DUTs (see TABLE II) have shown that the time base inaccuracy under all the test conditions affects the FE with an expected value of 801 μHz, but it can be reduced to 115 μHz after compensation.

TABLE II
STATISTICS OF THE PWM BLOCK

|  |  | $\mathbb{E}[\cdot]$ | $\sigma_{\hat{R}}$ | $\sigma_{R_X}$ | $\sigma_{R_T}$ |
|---|---|---|---|---|---|
| $e_{\bar{R}}$ (ppm) |  | -16.0 | 0.98 |  | 3.67 |
| Temperature (°C) | 0 | -19.9 |  | 2.72 |  |
|  | 10 | -19.0 |  | 2.68 |  |
|  | 20 | -17.3 |  | 2.47 |  |
|  | 30 | -14.2 |  | 1.93 |  |
|  | 40 | -13.2 |  | 1.42 |  |
|  | 50 | -12.5 |  | 1.40 |  |

As previously mentioned, the time base inaccuracy manifests also on the TVE in the form of a cyclostationary process, whose maximum magnitude can be evaluated over the synchronization interval of the time base with the PPS. A maximum TVE value equal to 0.5 % has been found at 1 s delay from the PPS over the entire temperature range, which can be reduced to 0.11 % after compensation. However, a more accurate compensation can be obtained if a temperature sensor is available on-board.

D. $\dot{\Lambda}_\tau$ – Error of the PLL block

The delay introduced by the PLL in restarting the time base after the PPS interrupt has occurred, has an intrinsic stochastic nature being the result of the OS process scheduling activity. Purpose of the characterization is to assess how the OS activity affects the synchronization delay $\tau$ by quantifying the effect of the underlying mechanisms. Therefore, besides the normal Idle state, the test is iterated under each of the following four stress conditions:

- CPU performs calculations of the function *sqrt*
- IO commits all scheduled data via low-level I/O system calls to non-volatile storage buffer using standard *sync* system calls
- HDD writes (delete) 1 MB block data on (from) the storage via standard *write/unlink* system calls
- VM dynamically allocates (free) 256 kB memory blocks via standard *malloc/free* system calls.

In addition, all test are performed using the same system image cloned on two SD cards with different speed class: a Class 4 and a Class 10 respectively with minimum sequential writing speed of 4 MB/s and 10 MB/s.

The synchronization delay is assessed via Two-Signal Edge-Separation method by measuring the time delay between the rising edges of the PPS on the Aux input and the first conversion pulse after synchronization on the Gate input of the DAQ board. The active edge on the Aux input triggers the internal counter, which counts the $N$ edges of the internal known time base $F_k$ until an active edge on the Gate input is detected. Similarly to (42), the delay is then calculated as

$$\tau = \frac{N}{F_k} = NT_k \quad \rightarrow \quad e_\tau = T_k. \qquad (47)$$

where the maximum error is equal to the time base period. In terms of phasor notation, the response of the PLL block is described by a pure random phase delay

$$\widehat{\Lambda}_\tau = e^{j\omega\hat{\tau}} \qquad (48)$$

where the statistics of $\hat{\tau}$ are calculated over 1000 measurements for each stress condition. Except for the VM stress test, all the obtained PDFs, whose Q-Q plot are reported in Fig. 4, show a normal distribution that extends over $3\sigma$ on the negative tail and $1\sigma$ on the positive one. The lower bound is evidently the result of the minimum time required to handle the interrupt, whereas the positive skew is related to the unpredictable OS activity.

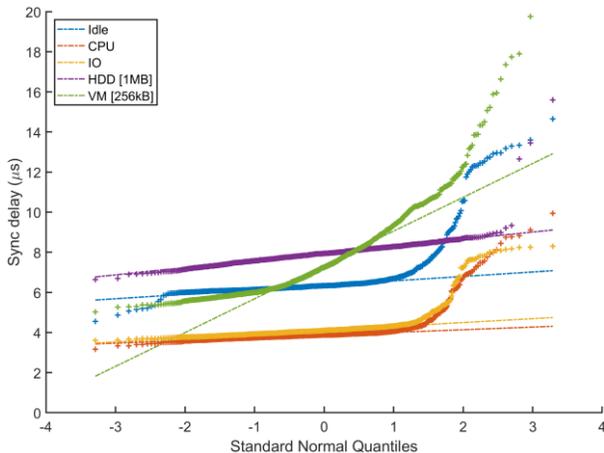

Fig. 4. Q-Q plot of the synchronization delay under different conditions

The only relevant difference observed in the test on different SD is – as expected – during the HDD stress test: an increment in the mean value of the delay of almost 10 % for the Class 4 SD. The most deleterious effect is reported in conjunction with the Virtual Memory stress test, suggesting to limit the number of dynamic operations on the memory. Conversely, the CPU and IO stress test report a beneficial effect with respect to the Idle state, suggesting to disable the CPU frequency scaling to obtain improved and more repeatable performance.

TABLE III
STATISTICS OF THE PLL BLOCK

| Test | Min | Max | Mean | Std$_{mean}$ | Mode | Std$_{mode}$ |
|---|---|---|---|---|---|---|
| Idle | 4.55 | 14.65 | 6.59 | 1.07 | 6.31 | 1.11 |
| CPU | 3.16 | 9.94 | 4.02 | 0.69 | 3.87 | 0.70 |
| IO | 3.60 | 8.30 | 4.25 | 0.67 | 4.12 | 0.68 |
| HDD | 6.64 | 15.59 | 7.95 | 0.49 | 7.93 | 0.50 |
| VM | 3.12 | 20.94 | 7.67 | 1.93 | 6.00 | 2.55 |

All values are in microseconds

Statistics results reported for sake of clarity in TABLE III, show a maximum delay introduced by the system under all conditions below 20 μs, which corresponds to a maximum TVE equal to 0.62 %. Results also show a minimum delay of 3.1 μs that can be considered to all effects as a systematic error affecting the TVE to the extent of 10 %, but alternative compensations can be taken into account by considering for instance the maximum measured mean value.

## V. RESULTS AND VALIDATION

The discussed error model is validated in an extended test session of the duration of 30 s. Over this period, the DAQ board is responsible for the synchronous generation of the analog and the PPS signals, while the DUT performs the acquisition of the same signal in parallel on the eight input channels. As result, the test generates 240 uncorrelated sub-sequences on which statistics are first calculated and then compared with the model.

First, both the phasor Z associated to the acquired signal in (13) and the phasor X of the reference 10 V amplitude signal are estimated over a sliding window of duration $T_p = 1/f$ by calculating the Fourier coefficient (49) of the first harmonic for a nominal frequency $f$ of 50 Hz, where $s(t)$ is the analyzed signal and $n$ is the harmonic order.

$$c_n(t) = \frac{2}{T_p} \int_t^{t+T_p} s(t) \cdot e^{-j2\pi nft} dt \qquad (49)$$

TABLE IV
SYSTEM RESPONSE MODEL PARAMETERS

| Symbol | Mean | Standard Uncertainty | Description | Unit |
|---|---|---|---|---|
| $e_P$ | -4429 | 255 | AAF phase error | μrad |
| $e_H$ | -9.81 | 1.13 | AAF gain error | ppm |
| $e_G$ | -4459 | 134 | ADC gain error | ppm |
| $e_R$ | -16.02 | 3.67 | PWM gain error | ppm |
| $e_\tau$ | -7.93 | 0.7 | PLL delay error | μs |

Subsequently, using the values obtained from the error characterization, summarized for sake of clarity in TABLE IV, the combined expected system response and the associated uncertainty are estimated in (50), where the maximum combined uncertainty must be evaluated by taking all the terms with the same sign.

$$\begin{aligned} e_{\overline{\Lambda}} &= e^{(e_H + e_G) + j(e_P + \omega t e_R + \omega e_\tau)} \\ u_\Lambda &= e^{(u_H \pm u_G) + j(u_P \pm \omega t u_R \pm \omega u_\tau)} \end{aligned} \quad (50)$$

Hence, the TVE associated to the acquired signals and its mean are computed together with the TVE expected from the model. The error obtained over the entire session interval is sliced between two consecutive PPS instants and shown in Fig. 5.

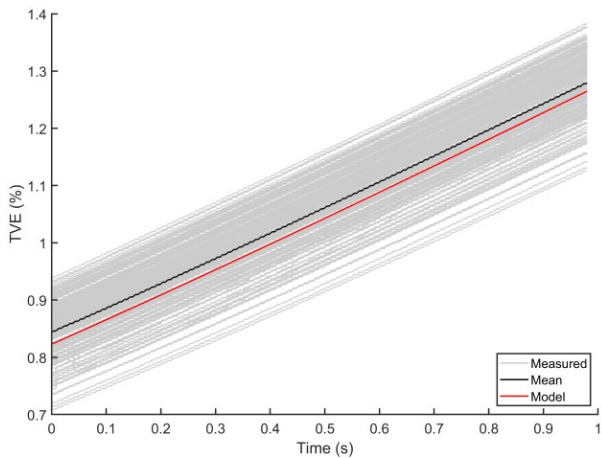

Fig. 5. Total Vector Error measured over 240 sequences (gray), mean value (black) and expected value from the model (red)

The TVE is then recalculated after applying the proposed compensation method to the entire sequence. Results reported in Fig. 6 show the mean value of the residual TVE after compensation, together with the associated Type A expanded uncertainty expressed at 99.9% confidence level using a coverage factor of $k = 3.3$. In addition, the Type B uncertainty predicted by the model with the same confidence interval is also reported for sake of comparison.

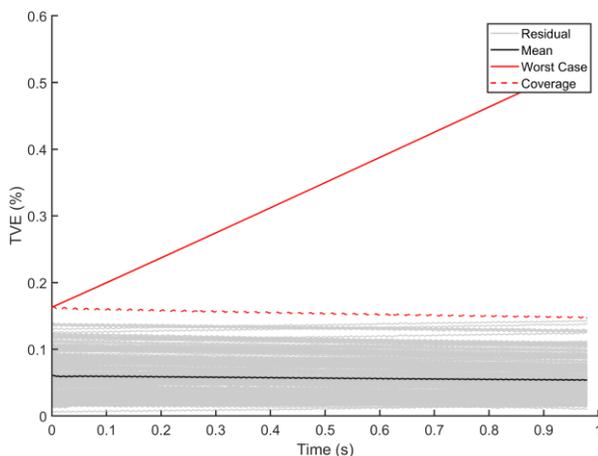

Fig. 6. Residual Total Vector Error after compensation (gray), mean value (black), worst case residual from the model (red solid) and confidence interval (red dashed)

The simulated error closely matches the measured one both qualitatively and quantitatively, resulting in a reduction of the average TVE of more than one order of magnitude. Furthermore, the simulated uncertainty equals the measured one when the PPS synchronization occurs, thus suggesting a proper modelling of the involved effects. Whereas its linear increment accounts for extreme thermal drifts of the local oscillator that might lead to a progressive cumulation of the phase shift between the reference and the measured phasor if not compensated with respect to the actual temperature.

The resulting 0.0556 % residual error after compensation can be further decomposed in 48 ppm magnitude error and 0.03° angle error, approximately half and three times the respective the 0.0001 per-unit voltage and 0.01° angle accuracy values reported by NASPI [23] for the ARPA-E μPMU. A similar consideration also applies to the accuracy predicted by the model: 134 ppm standard uncertainty on the magnitude and 0.09° on the angle. However, the increased angle uncertainty has to be ascribed to the uncompensated thermal drift of the local oscillator, which can be largely cancelled if a temperature compensation based on the values in TABLE II is implemented.

## VI. CONCLUSION

Errors in PMU architectures based on SBCs can be easily and effectively described by the proposed error model. Besides defining a small subset of measurable quantities that can be easily ascribed to well identified mechanisms, the model is able to both match the observed error with a high degree of accuracy and to provide a maximum boundary to the same. The residual average TVE of 0.055 %, more than one order of magnitude smaller than the maximum threshold fixed by the standard, is a very significant result if considered that was achieved on cheap and general-purpose platform based on a non-real-time OS. Furthermore, it can be further decreased after characterization of the sole PLL delay error for a specific implementation of the phasor estimation algorithm, for which the values in Table III should be used as guidelines for optimal code development.

The proposed model can be extended to those PMU architectures that rather make use of closed-loop PLL based on a local oscillator, however in that case it would not be possible to consider the time base deviation and the PPS synchronization delay as two uncorrelated quantities.

The model presented in this paper results also of great value for design purposes. In fact, it allows for a clear quantification of each single error contribution, which can be treated separately with ad-hoc design solutions aimed to reduce its impact.

ACKNOWLEDGMENT

This work was supported by SOGNO, a European project funded from the European Union's Horizon 2020 research and innovation programme under grant agreement No. 774613.